\documentclass[amsmath,amssymb,twocolumn]{revtex4-2}
\usepackage[T1]{fontenc}
\usepackage{graphicx}
\usepackage{subfig}
\usepackage{epsfig}
\usepackage{dcolumn}
\usepackage{multirow}
\usepackage{rotating}
\usepackage{verbatim}
\usepackage{bm}
\usepackage{color}
\usepackage{soul}
\usepackage{float}
\usepackage{hyperref}
\usepackage[utf8]{inputenc}

\def\lsim{\raise0.3ex\hbox{$<$\kern-0.75em\raise-1.1ex\hbox{$\sim$}}}
\def\gsim{\raise0.3ex\hbox{$>$\kern-0.75em\raise-1.1ex\hbox{$\sim$}}}

\def\beqa{\begin{eqnarray}}
\def\eeqa{\end{eqnarray}}

\begin{document}

\title{Role of nuclear gluon distribution on particle production in heavy ion collisions}
\author{L. S. Moriggi}
\email{lucas.moriggi@ufrgs.br}
\author{G. M. Peccini} 
\email{guilherme.peccini@ufrgs.br}
\author{M. V. T. Machado}
\email{magnus@if.ufrgs.br}
\affiliation{High Energy Physics Phenomenology Group, GFPAE. Institute of Physics, Federal University of Rio Grande do Sul (UFRGS)\\
Caixa Postal 15051, CEP 91501-970, Porto Alegre, RS, Brazil}

\begin{abstract}

The transverse momentum spectra of hadrons are calculated from the unintegrated gluon distribution (UGD) within the $k_T$-factorization framework at small $x$. Starting from $pp$ collisions, the modification caused by the nuclear medium is incorporated in the UGD at high energies, which is related to the nuclear shadowing phenomenon. Moreover, we consider that particle production from minijet decay is not enough to explain the $p_T$ spectra in $AA$ collisions due to collective phenomena that take place after the hard collision. The Boltzmann-Gibbs blast wave (BGBW) distribution is utilized in order to evaluate the distribution of particle production in equilibrium. Data from the ALICE Collaboration for $PbPb$ collisions at $\sqrt{s}=2.76$ TeV are analyzed and the nuclear modification factor for pion production is computed.

\end{abstract}

\maketitle

\section{Introduction}

The modification of nuclear structure functions at small Bjorken $x$, compared to those for free nucleon observed in DIS \cite{Aubert:1983xm} (i.e, EMC effect, antishadowing and shadowing), can be attributed to distinct phenomena \cite{Rith,Arneodo:1992wf,Malace_2014}. In particular, the shadowing seen in nuclear DIS \cite{Armesto_2006} for $x \lesssim 0.01$ is characterized by depletion of $F_2^A$ with respect to $F_2^N$. This effect has been explored within the Color Glass Condensate (CGC) saturation approach \cite{Gelis:2010nm}, which in turn predicts a saturation scale, $Q_s(x)$. Such a quantity establishes the region where the increasing of the unintegrated gluon distribution (UGD) on $x$ is tamed. It is expected that for heavy ion collisions the saturation scale is enhanced compared to the nucleon case  and it is deeply connected to shadowing corrections. Hence, the nuclear saturation scale characterizes the dense system. In $pA$ collisions, the depletion of cross section related to the $pp$ mode is also evident in the region of small $p_T$ and may be well described in the context of saturation formalism \cite{Lappi:2013zma,DHJ,Magno}. However, in $AA$ collisions the particle production presents more complex behavior which cannot be explained solely by nuclear effects in the  gluon distribution at the initial state. While the collinear factorization mechanism is well established at large $Q^2$, the saturation formalism or CGC makes use of $k_T$ factorization in the small $x$ domain in order to describe minijet production of gluons where the UGD, $\phi (x,k_T)$, depends on the transverse momentum $k_T$. This distribution is related to the dipole cross section, $\sigma_{dp}(x,r)$, the latter being directly extracted from DIS at small $x$. This formalism has been employed to describe the $p_T$ spectra of produced hadrons at RHIC and LHC \cite{Armesto:2004ud,ALbacete:2010ad,Kharzeev:2001gp,Tribedy:2011aa,Lappi_2011,Levin:2011hr,Albacete:2010bs,Duraes:2016yyg}. 

In our previous work \cite{Moriggi_2020}, we have computed the $p_T$ spectra of produced hadrons in $pp$ collisions over a large interval of the scaling variable, $\tau=p^2_T/Q^2_s(\sqrt{s})$. In addition, a phenomenological parametrization for an UGD was proposed, which has a power-like behavior at high $k_T$. As a result, it was shown that scaling is a good approximation for describing the spectra at different collision energies, $\sqrt{s}$. Regarding the nuclear case, the gluon distribution should be properly modified in order to include cold matter effects.

In the collinear factorization framework, the nuclear parton distribution functions (NPDFs) may be obtained through the parametrization of the distribution at $Q^2=Q^2_0$. Then, these functions are evolved toward higher $Q^2$ by means of Dokshitzer-Gribov-Lipatov-Altarelli-Parisi (DGLAP) evolution equations \cite{Kovarik:2015cma,EPPS16,NNPDFnuclear}. Alternative approaches concerning collinear NPDFs include the consideration of impact parameter dependence \cite{Helenius:2012wd,Shao:2020acd}, nonlinear effects \cite{WANG20171}, or dependence of microscopic effects on nuclear modification as performed in Ref. \cite{Kulagin:2014vsa}. Another successful approach is the leading twist theory of nuclear shadowing \cite{Frankfurt:1998ym,Frankfurt:2003zd,Guzey:2009jr}, which is based on generalization of the
Gribov-Glauber theory for nuclear shadowing. Such a formalism is able to predict next-to-leading
order (NLO) NPDFs and diffractive NPDFs as a function of the impact parameter and nuclear generalized parton distributions as well (see a review in \cite{Frankfurt:2011cs} and references therein). 
 The determination of unintegrated distributions can also be carried out from the integrated distribution through the resummation of soft emissions by the Sudakov form factor \cite{Kimber:1999xc,Watt:2003mx} that may be applied in the  nuclear case \cite{deOliveira:2013oma,Modarres:2019ndk}. In the approach of Ref. \cite{Kutak:2012rf}, the UGD can be derived from an unified formalism of BK and DGLAP equations taking into account saturation effects. In Ref. \cite{Bury:2017xwd}, it has been shown that such nonlinear effects might yield important modifications on forward jet production. Partonic distributions depending on transverse momentum were also obtained from the parton branching approach for the proton \cite{Hautmann:2017xtx,Martinez:2018jxt} and for the nuclear case, as well \cite{Blanco:2019qbm}. In the context of saturation/CGC framework, distinct ways of obtaining the nuclear UGD were proposed \cite{Pirner:2001an,Armesto:2002ny,Kharzeev:2001yq,Kharzeev:2004if,Drescher:2006ca,Albacete:2012xq,Betemps:2009da}. Some of the approaches discussed above has been implemented in Monte Carlo generators. A well-known example is the {\sc HIJING++} heavy-ion Monte Carlo \cite{Papp:2018qrc,Biro:2018ntr,Biro:2019ijx}, which is an event generator for parton and particle production in high-energy hadronic and nuclear collisions based on QCD-inspired models for multiple jet production. It incorporates mechanisms such as multiple minijet production, soft excitation, nuclear shadowing of PDFs, and jet interactions in dense hadronic matter \cite{Wang:1991hta,Deng:2010mv}.
 
 In the present work, the nuclear effects are incorporated in the nuclear UGD by means of the Glauber multiple scattering theory as performed in Refs. \cite{Tribedy:2011aa,Lappi_2011}. In $pp$ collisions, particle production can be obtained within the $k_T$ factorization assuming the local hadron-parton-duality (LHPD). The final hadron spectra are directly related to those ones from produced gluons at the initial state in a good approximation \cite{Moriggi_2020,Levin:Rezaeian}. As already  pointed out, such scenario might not be appropriate for collisions of heavy ions, where there are collective effects that modify the $p_T$ spectra of produced hadrons with respect to the initial state. However, it has been argued in Refs. \cite{Tripathy_2016,Tripathy:2017kwb,Qiao_2020}
that $p_T$ spectra can be well described by making a time separation in the relaxation time approximation (RTA) formalism of the Boltzmann transport equation \cite{Florkowski:2016qig} among produced hadrons at the initial hard collision (which is parametrized by Tsallis \cite{Tsallis:1987eu} or Hagedorn distributions \cite{Hagedorn}) and produced hadrons by the system in thermal equilibrium. The collective radial flux plays an important role on the distribution form. Moreover, models with two components are successful over a wide amount of data concerning particle production at high collision energies \cite{bylinkin2013hadroproduction,Bylinkin_2016,Giannini:2020gys}. The hadronic spectra is decomposed into two parts,the first one being related to Boltzmann statistics and the second one based on power law behavior that captures the aspects of perturbative QCD predictions.

In this work, we propose a nuclear unintegrated gluon distribution that embeds the shadowing verified in DIS. The cross section for minijet production, which is driven by gluons within the $k_T$-factorization framework, is obtained. Additionally, the effects caused by the medium at the final state are incorporated by the formalism of the Boltzmann equation in the relaxation time approximation. The produced particles at the initial hard collision are calculated by using the $k_T$ factorization,  whereas the $p_T$ spectra due to hydrodynamics expansion is given by Boltzmann-Gibbs blast wave (BGBW) distribution. The  distribution parameters are determined from data of pion production at different centralities as measured by ALICE for $\sqrt{s}=2.76$ TeV \cite{Abelev:2013vea}, and the nuclear modification factor,$R_{AA}$, is predicted. A similar analysis has been performed in Ref. \cite{Qiao_2020} without taking into account nuclear shadowing. In that sense, the aim here is to understand the impact caused by the modification of gluon distribution at small $x$ on the observed nuclear modification factor.

This paper is organized as follows. In the next section, we present the details on the determination of the nuclear UGD, $\phi_A(x,k_T)$,  from the free nucleon distribution, $\phi_p(x,k_T)$. This is achieved using the multiple scattering formalism as well as the predictions of the spectra of produced hadrons using the $k_T$ factorization with these modifications. We also describe the hadron production at the final state from the BGBW distribution and constrain the relevant parameters. In Sec. III, predictions are compared against data for pion production and nuclear modification factor. An analysis on the interpretation of the obtained parameters is performed. Finally, in Sec. IV we outline the main results and present  conclusions.

\section{Theoretical framework and main predictions}
\label{sec:model}
\begin{table*}[t]
\centering
\caption{Values of fitted parameters in each centrality class for production of charged pions in $PbPb$ collisions at $\sqrt{s}=2.76$ TeV.}

\label{tab:pars}
\begin{tabular}{|c|c|c|c|c|c|c|c|}
Centrality(\%)       & $t_f/t_r$  & $T$ (GeV)      & $\left<\beta\right>$   &   $\frac{\chi^2}{\mathrm{dof}}$ \\\hline
0-5   & $2.125\pm0.119$ & $0.1110\pm0.0132$ & $0.5740\pm0.0236$ & 0.766 \\ \hline
5-10  & $1.962\pm0.035$ & $0.1094\pm0.0130$ & $0.5781\pm0.0227$ & 0.680 \\ \hline
10-20 & $1.770\pm0.029$ & $0.1119\pm0.0144$ & $0.5742\pm0.0256$ & 0.636 \\ \hline
20-40 & $1.417\pm0.028$ & $0.1023\pm0.0164$ & $0.5905\pm0.0265$ & 0.559 \\ \hline
40-60 & $0.970\pm0.036$ & $0.0781\pm0.0217$ & $0.6217\pm0.0270$ & 0.407 \\ \hline
60-80 & $0.621\pm0.029$ & $0.0498\pm0.0053$ & $0.6469\pm0.0046$ & 0.216 \\ \hline
\end{tabular}
\end{table*}
The nuclear UGD may be obtained from the nucleon distribution by using the Glauber-Mueller \cite{Glauber:1955qq,Mueller:1989st} approach for multiple scattering. It has been carried out, for instance, in Ref. \cite{Armesto:2002ny}. In this case, the dipole scattering matrix in configuration space, $r$, can be determined from the cross section for dipole scattering off a proton,
\begin{eqnarray}
\label{eq:SdA}
S_{dA}(x,r,b)=e^{-\frac{1}{2} T_A(b)\sigma_{dp}(x,r)},   
\end{eqnarray}
where $T_A(b)$ is the thickness function which depends on the impact parameter $b$. In the present work, a Woods-Saxon-like parametrization for the nuclear density \cite{DEVRIES1987495} with normalization $\int d^2b T_A(b) = A $ has been applied for a lead nucleus.  The related nuclear UGD is given by
\begin{equation}
\label{eq:fiA}
\phi_A(x,k_T^2,b)=\frac{3}{4\pi^2\alpha_s}k_T^2\nabla^2_k \mathcal{H}_0 \left\{ \frac{1-S_{dA}(x,r,b)}{r^2}\right\} \ ,
\end{equation}
$\mathcal{H}_0 \left\{ f(r) \right\}=\int rdr J_0(k_Tr)f(r)$ being the Hankel transform of order zero.  

For the proton case, a homogeneous target with radius $R_p$ is considered so that the dependence on impact parameter is factorized as $S_{dp}(x,r,b)=S_{dp}(x,r)\Theta (R_p-b)$. In the limit of large dipoles, $S_{dp}(x,r) \rightarrow 0$, and the cross section reaches a maximum, $\sigma_0=2\pi R_p^2$. Within the parton saturation formalism, the gluon distribution should have a maximum around  $k_T=Q_s(x)$. One of the features of this formalism is the presence of geometric scaling in the observables, which become dependent on the ratio $Q^2/Q_s(x)$ rather than of $Q^2$ and $x$ separately. 

It has been proposed in Ref. \cite{Moriggi_2020} a gluon distribution based on geometric scaling of high $p_T$ spectra of produced hadrons in $pp$ collisions: 
\begin{eqnarray}
\label{eq:UGDp}
\phi_p(\tau)=\frac{3\sigma_0 (1+\delta n)}{4\pi^2\alpha_s}\frac{\tau}{ \left(1+\tau \right )^{(2+\delta n)}} \ ,
\end{eqnarray}
where the scaling variable is defined as $\tau=k_T^2/Q_s^2(x)$ and the parameter $\delta n$ establishes the powerlike behavior of the spectra of produced gluons at high momentum $\tau \gg 1$. The cross section for dipole scattering in coordinate space, $r$, may be written as 
\begin{eqnarray}
\label{eq:DISc}
\sigma_{dp}(\tau_r)=\sigma_0\left (  1-\frac{2(\frac{\tau_r}{2})^{\xi}K_{\xi}(\tau_r)}{\Gamma(\xi)} \right ),
\end{eqnarray}
in which $\tau_r = rQ_s(x)$ is the scaling variable in the position space and $\xi =1+\delta n$. Therefore, by placing $\sigma_{dp}(x,r)$ in Eq. (\ref{eq:SdA}) the nuclear gluon distribution is obtained directly from Eq.  (\ref{eq:fiA}).

The spectra of produced gluons in the initial hard collision, given an impact parameter $b$, can be calculated in the $k_T$-factorization formalism \cite{Gribov:1983fc}:
\begin{eqnarray}
\label{eq:fatkt}
E\frac{d^3 N(b) }{dp^3}^{AB \rightarrow g+X}&=&\frac{2\alpha_s}{C_F}\frac{1}{p_T^2}\int d^2s \, d^2k_T\,  \phi_A(x_A,k_T^2,s) \nonumber \\
&\times & \phi_B(x_B , (p_T-k_T)^2, b-s).
\end{eqnarray}
Above, $p_T$ is the transverse momentum of the produced gluon and $x_A$ and $x_B$ are the gluon momentum functions in the nucleus $A$ and $B$, respectively. They are expressed in terms of the rapidity $y$ in the following way:
\begin{eqnarray}
x_A=\frac{p_T}{\sqrt{s}}e^y, \qquad x_B=\frac{p_T}{\sqrt{s}}e^{-y}. 
\end{eqnarray} 

In the LHPD approximation, the spectra of produced hadrons is directly related to the minijet originated in gluons. In that case, we consider the hadron being produced with momentum $p_{Th}=\left<z\right>p_T$:
\begin{eqnarray}
\label{eq:hadron2}
\frac{d^3N^{AB\rightarrow h}}{d^2p_{Th}dy}=\frac{K}{\left< z\right>^2}\frac{d^3N^{AB\rightarrow g}}{d^2p_{Th}dy}\left(p_T=\frac{p_{Th}}{\left< z\right>}\right).
\end{eqnarray}
The parameters $K$ and $\left<z\right>$ are the same as those obtained for the spectra $pp\rightarrow \pi^0+X$. Also, it is important to notice that Eq. (\ref{eq:fatkt}) diverges for $p_T \rightarrow 0$, so that one needs to apply a cut associated with the jet mass $p_T^2\rightarrow p_T^2+m_j^2$, where the value $m_j=0.2$ GeV is used.

The suppression of hadron production in nuclear collisions due to the nuclear modifications on gluon distribution may be quantified through the following ratio:

\begin{equation}
\label{eq:RAAs}
R^{shadow}_{AA}(b)=\frac{ \frac{dN_{AA}(b)}{d^2p_{Th}dy}}{ \int d^2s \, T_A(s)T_B(b-s)\frac{d\sigma_{pp}}{d^2p_{Th}dy}}.
\end{equation}

It can be seen that for small dipoles, i.e., $r \rightarrow 0$ (or, equivalently, for high $k_T$), it is possible to expand Eq. (\ref{eq:SdA}), which leads to $S_{dA} \sim T_A(b)\sigma_0 S_{dp}(x,r)$ and $R^{shadow}_{AA}(b) \rightarrow 1$ for any value of $b$.
This scenario is strictly valid for the case of particle production from minijet yield that is originated in the initial hard interaction. On the other hand, in heavy ion collisions the initial hard scattering is followed by the formation of quark-gluon plasma (QGP), and the evolution of the system until the freeze-out in the hadronic phase has an important effect on final spectra. In Ref. \cite{Tripathy_2016}, it was proposed that the evolution of particle distribution due to their interaction with the medium is set up by the Boltzmann transport equation within the RTA:

\begin{equation}
f_{fin}=f_{eq}+(f_{in}-f_{eq})e^{-t_f/t_r},  
\end{equation}
$t_r$ being the relaxation time and $t_f$ the time of freeze-out. Thereby, the hard initial distribution ($t=0$) evolves until the final distribution $f_{fin}$ at $t=t_f$. The equilibrium distribution, $f_{eq}$, is characterized by the equilibrium temperature, $T$, and the relaxation time, $t_r$, the latter being responsible for determining the amount of time until the system reaches equilibrium. Following Ref. \cite{Qiao_2020}, we have considered that particle distribution in equilibrium can be evaluated by the BGBW model \cite{Schnedermann:1993ws}, 

\begin{equation}
\label{eq:BGBW}
f_{eq}\propto m_T\int_0^R rdrK_1\left( \frac{m_T\cosh (\rho))}{T}  \right )I_0\left( \frac{p_T\sinh (\rho))}{T}  \right ),
\end{equation}
$I_0$ and $K_1$ being the Bessel functions of first and second kinds of order zero and one, respectively. The quantity $m_T$ is the transverse mass, $m_T=\sqrt{p^2_{Th}+m_h^2}$, and the velocity profile $\rho$ is given by

\begin{equation}
\label{eq:rho}
\rho=\tanh^{-1}\left[ \left( \frac{r}{R}\right )^m \beta_s \right],
\end{equation}
where $\beta_s$ is the maximum velocity expansion of the surface with average transverse velocity $\left< \beta \right>=\frac{2}{2+m}\beta_s$.

The Tsallis distribution has been used in Refs. \cite{Tripathy_2016,Tripathy:2017kwb,Tripathy:2017nmo,Qiao_2020,Younus:2018mrk} in order to constrain the initial distribution without taking into account nuclear shadowing. That distribution can be deeply understood in the context of the fractal structures present in QCD or in general Yang-Mills theories \cite{Deppman:2017fkq,Deppman:2019yno}. It implies the need of Tsallis statistics (TS) to describe the thermodynamics aspects of the fields, and the entropic index of the TS can be obtained in terms of the field fundamental parameters (see Ref. \cite{Deppman:2020jzl} for a recent review). In our approach, we have shown that one can describe pion and charged hadrons distributions through the $k_T$ factorization framework. By utilizing the UGD of Eq. (\ref{eq:UGDp}), such a formalism produces a distribution with similar features of Tsallis distribution, which for $\tau \geq 1$ can be approximated (neglecting the nuclear effects) by

\begin{equation}
f_{in} \sim \frac{\xi}{\xi -1}\left( 1-\frac{1+\xi\tau}{(1+\tau)^{\xi}}\right )\frac{1}{( 1+\tau)^{1+\xi}}.
\end{equation}

The nuclear modification is introduced into the nuclear UGD, and the hard initial distribution, $f_{in}$, is taken from Eqs. (\ref{eq:fatkt}) and (\ref{eq:hadron2}). The saturation scale and the power index, $\delta n$, were parametrized in the following way \cite{Moriggi_2020}:

\begin{eqnarray}
\label{eq:pars}
 \delta n (\tau ) &=& a \tau ^b ,\\
 Q_s^2(x)&=& \left( \frac{x_0}{x}\right) ^{0.33}, 
\end{eqnarray}
in which the parameters $a$, $b$ and $x_0$ were obtained by fitting the HERA data (see Ref. \cite{Moriggi_2020} for details). Given these considerations, the hadron production in nuclear collisions is expressed as the following sum:

\begin{equation}
\label{eq:NAA}
 E\frac{d^3N_{AA}}{dp_h^3}=e^{-t_f/t_r}f_{in}(p_{Th})+(1-e^{-t_f/t_r})f_{eq}(p_{Th}).
\end{equation}

The distribution of produced particles in thermal equilibrium is given by Eq. (\ref{eq:BGBW}) and the nuclear modification factor for each centrality class reads as
\begin{equation}
\label{eq:RAA}
R_{AA}=\frac{\frac{d^3N_{AA}}{dp^3}}{\langle T_{AB}\rangle \frac{d^3\sigma_{pp}}{dp^3}},
\end{equation}
where $\langle T_{AB} \rangle $ is the mean value of nuclear overlap for a given centrality. 

\begin{figure}[t]
\includegraphics[width=\linewidth]{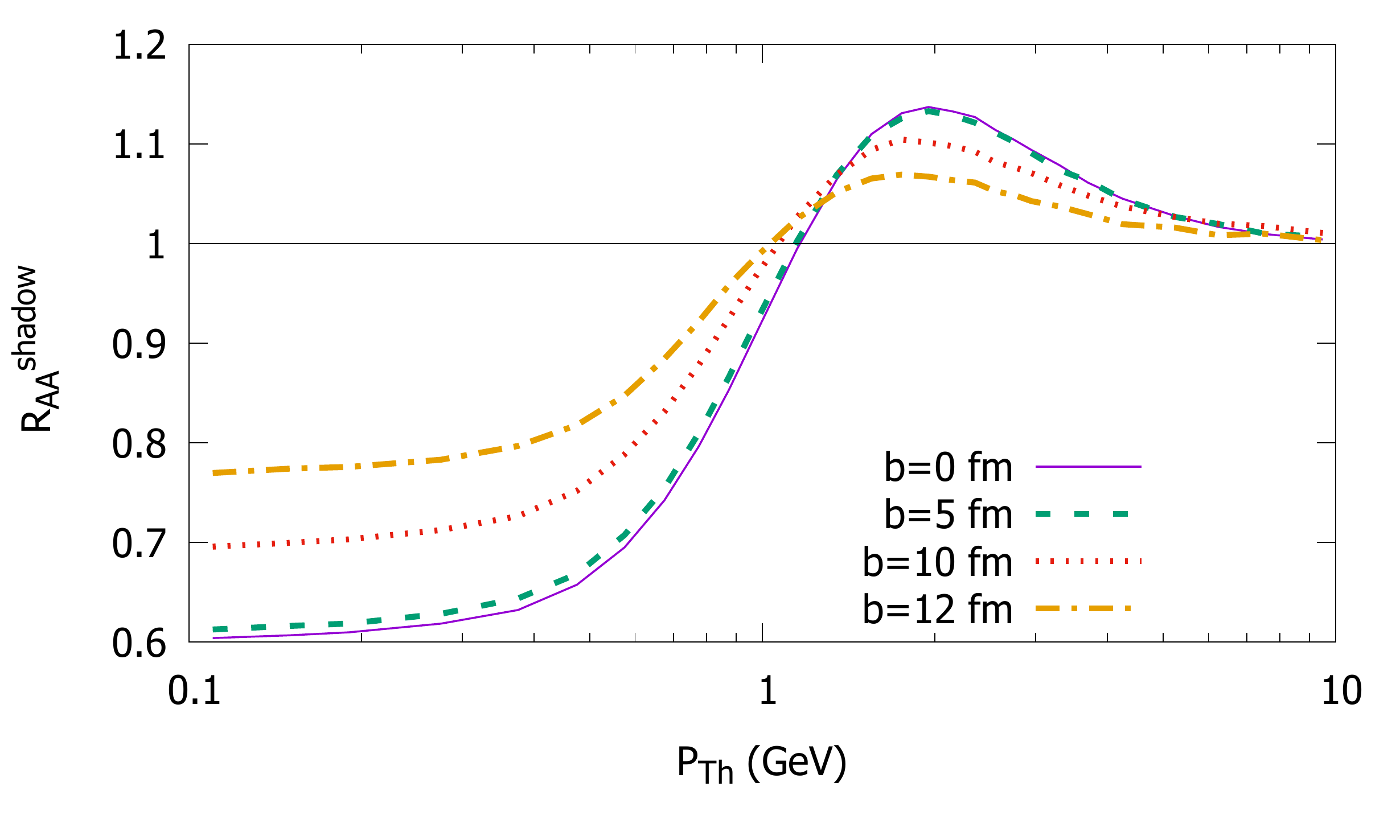}
\caption{Nuclear modification factor (defined in Eq. \ref{eq:RAAs}) calculated for different values of impact parameter $b$.} 
\label{fig:RAAshadow} 
\end{figure}

It is relevant to stress out that this definition is the same of that one utilized experimentally for the determination of $R_{AA}$ in each centrality class. Furthermore, both the $pp$ and $AA$ spectra are calculated from the same model described before \cite{Moriggi_2020}. Albeit the parametrization for the $pp$ cross section has been made for the sum of charged hadrons and neutral pions, in nuclear collisions we have restricted the analysis for pion spectra. The reason is that  proton production has strong influence on the region of middle $p_T$ spectra in $AA$ collisions\cite{Adcox:2001mf,Adcox_2005}. Such a phenomenon is known a baryon anomaly and other mechanisms are needed in order to explain it.

\begin{figure*}[t]
\includegraphics[width=0.7\linewidth]{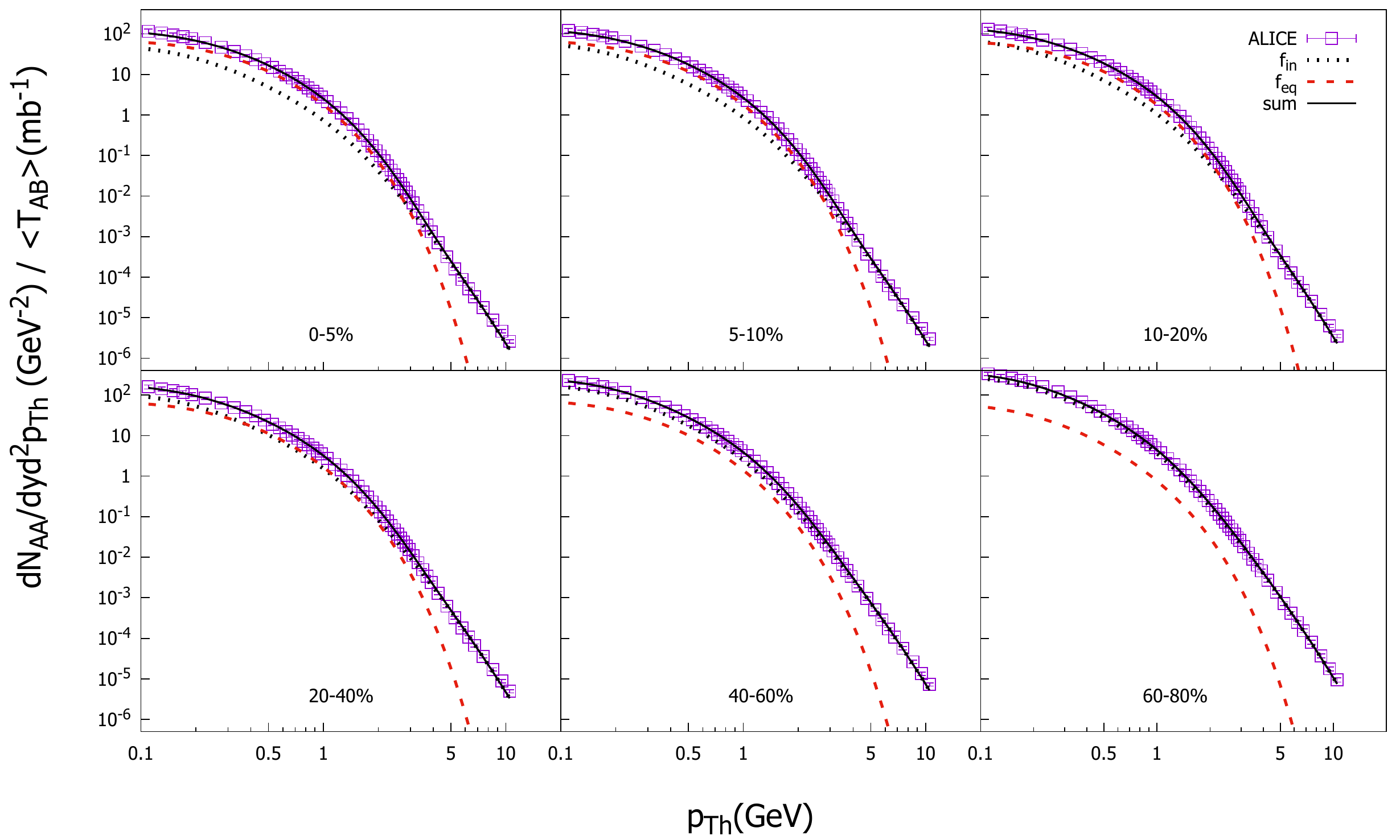}
\caption{Comparison between the final spectra defined by Eq. (\ref{eq:NAA}) with data from ALICE \cite{Abelev:2013vea}. The dotted and dashed lines denote the $f_{in}$ and $f_{eq}$ contributions, respectively.} 
\label{fig:NAA} 
\end{figure*}

\section{Results and discussion}
\label{sec:results}

Our analysis on $p_T$ spectra is limited by the kinematic window determined by  geometric scaling compatible with that observed at HERA \cite{Stasto,Qgs,BUW,Praszalowicz:2015vba}. For $\sqrt{s}=2.76$ TeV, one should have $p_{Th} \lesssim 10 $ GeV. First, the results of the nuclear modification factor are presented including only shadowing effect for different values of $b$ in Eq. (\ref{eq:RAAs}). In Fig. \ref{fig:RAAshadow}, it can be seen that $R^{shadow}_{PbPb}$ considerably enhances until a maximum point around $p_{Th}\sim 2$ GeV, which is the well-known Cronin peak \cite{Cronin:1974zm}. This is a result of multiple scattering, and the ratio further decreases until the limit $R_{AA}=1$ at high $p_{Th}$. The position of this peak is set by the saturation scale, $Q_s(x)$, and by the mean value of the gluon momentum fraction carried by the hadron, since the $pp$ cross section depends only on the scaling variable $\tau_h=\tau  \langle z\rangle^{2.33}$. In such a case we use $\left<z\right>=0.345$, which is the fitted value for $\pi^0$ spectra in $pp$ collisions for distinct values of $\sqrt{s}$. Higher values of $z$ should lead to the shift of this peak toward higher $p_T$. In more central collisions, the shape of $R_{AA}$ has little dependence on the impact parameter, whereas for more peripheral collisions the nuclear effects are weaker. Other models based on geometric scaling were proposed in order to determine the nuclear shadowing within the dipole approach (as done in Ref. \cite{Armesto:2004ud}, for instance). However, we did not get good results when utilizing this picture because the resulting modification factor grows rapidly with $p_T$. This issue has already been discussed in Ref. \cite{Baier:2003hr}. Furthermore, it is shown in Ref. \cite{Lappi:2013zma} that the multiple scattering formalism produces good results for the nuclear modification factor in $pA$ collisions.

\begin{figure*}[t]
\includegraphics[width=0.7\linewidth]{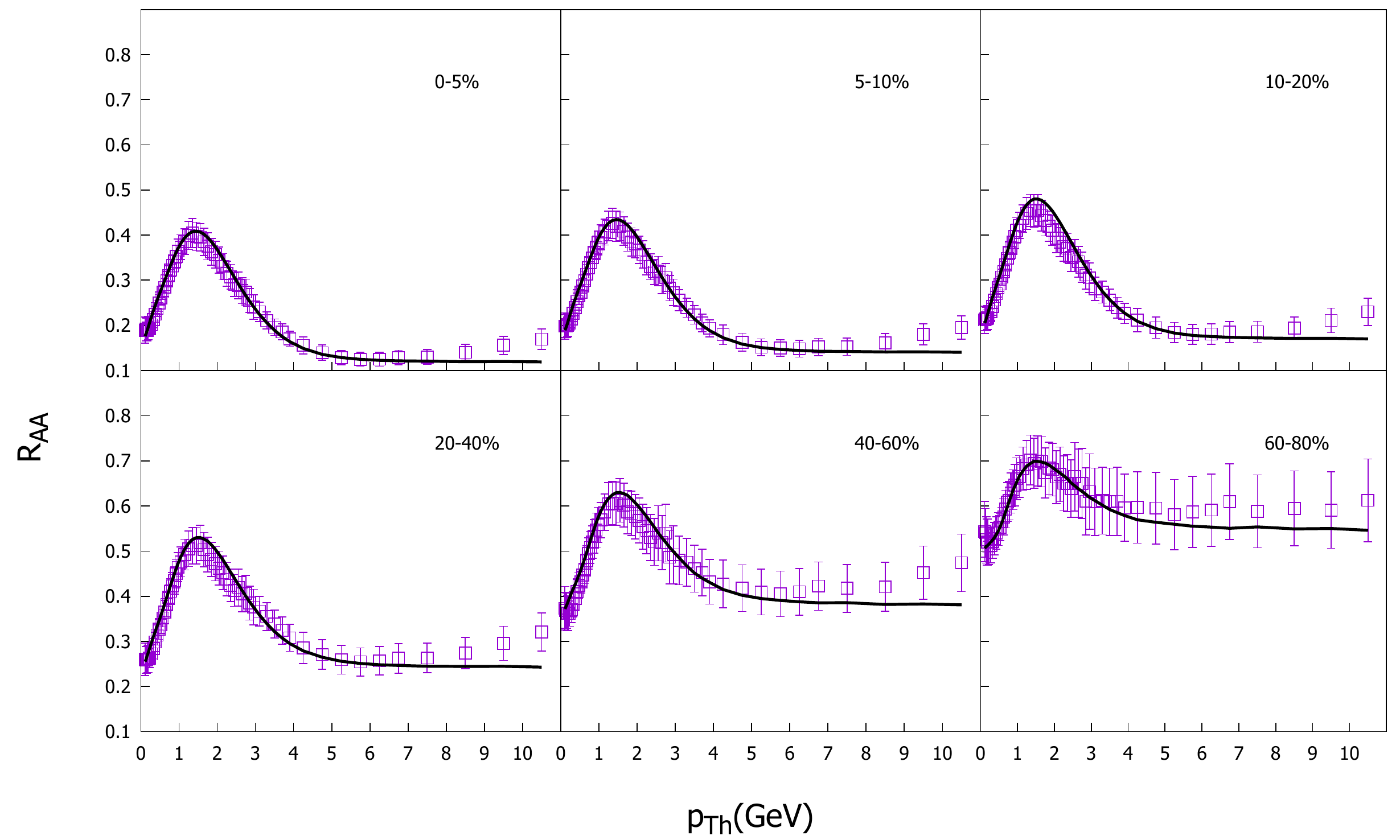}
\caption{Comparison between the nuclear modification factor obtained from Eq. (\ref{eq:RAA}) with data from ALICE \cite{Adam:2015kca} for different centrality classes.}
\label{fig:RAA} 
\end{figure*}

The determination of hadron final distribution, Eq. (\ref{eq:NAA}), is carried out by fitting the parameters of BGBW distribution in Eq. (\ref{eq:BGBW}). It has been considered a linear expansion profile, that is, $m=1$ in Eq. (\ref{eq:rho}), and the parameters $t_f/t_r$, $T$, and  $\beta_s$ are taken from data of $\pi^{+-}$ spectra at $\sqrt{s}=2.76$ TeV for each centrality class. The fit results are presented in Table \ref{tab:pars}. The values obtained for $t_f/t_r$ are very close to those ones in Ref. \cite{Qiao_2020}. Such a quantity lowers in terms of the centrality, indicating a higher relaxation time in peripheral collisions. This fact was understood as a result of initial distribution closer to equilibrium in more central collisions. 

Figure \ref{fig:NAA} displays the resulting curve  compared with data of $p_T$ spectra from $\pi^{+-}$ in Ref. \cite{Abelev:2013vea}. The dotted and dashed lines represent the two contributions in Eq. (\ref{eq:NAA}), namely, the hard initial distribution and the distribution of produced particles in equilibrium, respectively. It can be realized that for more central collisions (up to $10-20\%$) the region of small $p_T$ is dominated by BGBW-like thermal spectra, while for $p_T\gtrsim 4$ the leading mechanism is that one of produced particles in the hard initial collision through minijet yield initiated by gluons. For larger centralities, the contribution $f_{eq}$ becomes smaller even in the region of small $p_T$, and then the nuclear effects are basically determined by the shadowing of gluon distribution in the initial state.

Figure \ref{fig:RAA} shows the nuclear modification factor. The $pp$ cross section has been calculated in Ref. \cite{Moriggi_2020}. For the sake of consistency, the values of $\left< T_{AB}\right>$ that were employed are the same as in Ref. \cite{Adam:2015kca} for the determination of the experimental nuclear modification. In more central collisions, the modification on spectra is more intense than what is expected, which is explained by the shadowing seen in Fig. \ref{fig:RAAshadow}. It indicates that in this case particle production is caused by a distinct mechanism. In more peripheral collisions, the decreasing of cross section in $PbPb$ collisions with respect to the $pp$ one at small $p_{Th}$ is smaller and more compatible with what is expected if one considers only the shadowing of gluon distribution. The usual treatment for $R_{AA}$ is  based on pQCD, in which effects of energy
loss are absorbed in medium-modified parton fragmentation function in a dynamically expanding medium.  There are several prescriptions on how to include radiative
energy loss, and some of them also introduce a collisional contribution \cite{Djordjevic:2016vfo,Bianchi:2017wpt,Andres:2016iys,Chien:2015vja}. Our predictions agree with these theoretical approaches at large $p_T$ and describe correctly the peak at low $p_T$.

\begin{figure}[t]
\includegraphics[width=\linewidth]{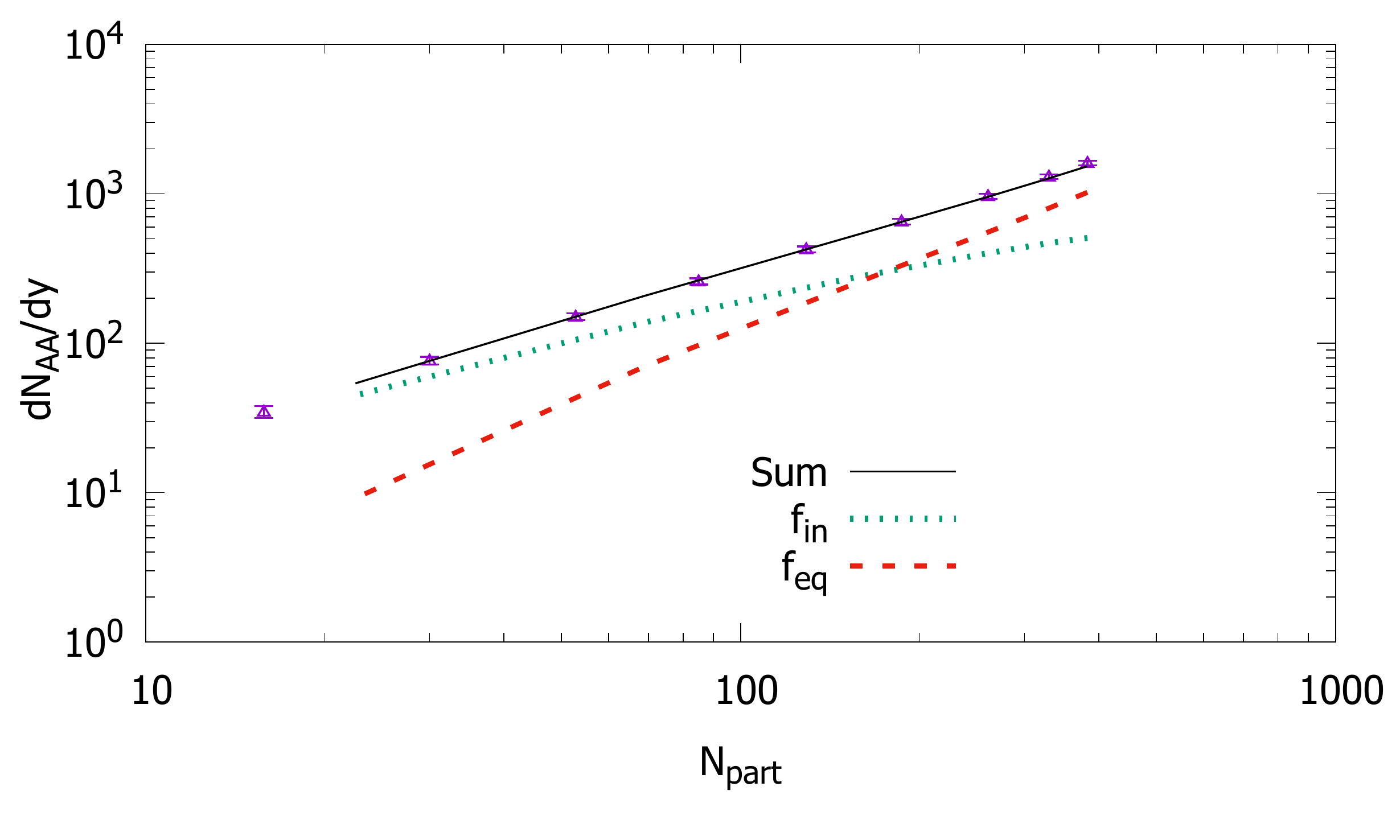}
\caption{Multiplicity of charged hadrons at $y=0$ compared with our predictions as a function of the number of participants, $N_{part}$. The dashed and dotted lines represent the $f_{in}$ and $f_{eq}$ contributions, respectively, in Eq. (\ref{eq:NAA}).}
\label{fig:dsigdy} 
\end{figure}

The inclusive multiplicity  of charged hadrons $dN/dy$ was also calculated by integrating 
the spectra over $p^2_{Th}$. In order to compare the pion distribution with data of $dN/dy$ at $y=0$ for charged hadrons, we have taken into account a correction of $5\%$ relative to the contributions of charged kaons and protons. Figure \ref{fig:dsigdy} presents our results compared against data from ALICE \cite{Aamodt_2011} as a function of the participant number, $N_{part}$. The line is an interpolation among the results from each centrality class. Interestingly, one may note that $f_{in}$ has a slower increase in terms of $N_{part}$, whereas $f_{eq}$ grows rapidly in the central region.

\begin{figure}[t]
\includegraphics[width=\linewidth]{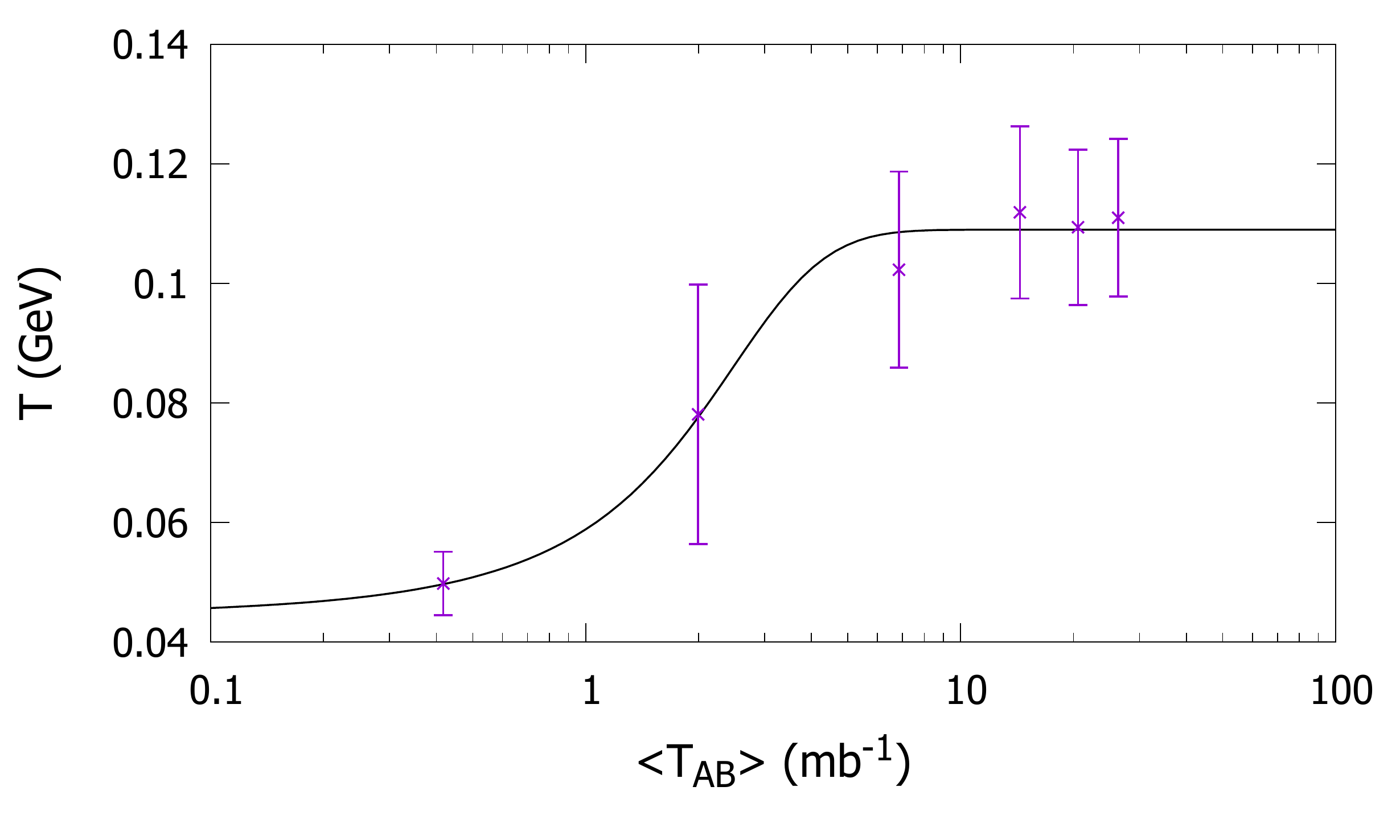}
\caption{Values of temperature of equilibrium resulting from fits utilizing the BGBW distribution in Eq. (\ref{eq:BGBW}) as a function of $\langle T_{AB}\rangle $. The line represents an interpolation and is discussed in the text.}
\label{fig:TBGBW} 
\end{figure}

Figure \ref{fig:TBGBW} displays the relation between the temperature and the nuclear overlap for each centrality. While the values obtained for $T$ and $\left<\beta\right>$ are close to those ones taken from fits using only the BGBW model as in Ref. \cite{Abelev:2013vea} for central collisions, (i.e., $T\sim 0.1$ GeV and $\left<\beta\right> \sim 0.6 $ are practically constants) in peripheral collisions the values of $T$ tend to be lower. Besides, since $T$ and $\left<\beta\right>$ are anticorrelated, we have an increasing of $\left<\beta\right>$ in more peripheral collisions. These differences occur due to the fact that in more central collisions the second term in Eq. 
(\ref{eq:NAA}) is the leading one at small $p_{Th}$, whereas for more peripheral collisions its contribution is much smaller. This effect can be interpreted in the following way: The first one is that the nuclear modification can be explained with good approximation only by the effect of nuclear shadowing. In this case, the fits of BGBW distribution parameters have large uncertainties. Instead, if one assigns a physical meaning for the temperature decreasing that  occurs in more peripheral collisions, the possible interpretation is that collective expansion that takes place in  there is similar to what happens in more central collisions for smaller collisions energies in which $T$ is significantly lower. This fact may occur since the mean number of collisions, $N_{coll}=\sigma_{in}(\sqrt{s}) \langle T_{AB}\rangle$, grows for more central collisions and higher energies. In Ref. \cite{bylinkin2013hadroproduction} it is pointed out that the temperature obtained from fits utilizing a Boltzmann-like distribution enhances in terms of the energy initial density (it depends on $N_{part}$ and $\sqrt{s}$) until it reaches a bound, and this was understood as a QGP phase transition temperature for hadrons. In Ref. \cite{Andronic:2005yp} it is shown that the dependence of $T$ in terms of $\sqrt{s}$ can be parametrized as $T=T_{lim}\left( 1- \frac{1}{A+Be^x} \right )$. This form can be applied for our case. Figure \ref{fig:TBGBW} presents the fitted line for this function with respect to $\langle T_{AB}\rangle$.

\section{Summary and conclusions}
\label{sec:conc}

In this work, we investigated the role of nuclear shadowing incorporated in the  gluon distribution through the saturation/CGC formalism applied to the spectra of produced gluons in heavy ion collisions at high energies. Through the Boltzmann equation formalism within the relaxation time approximation, the contributions from the hard initial distribution and from particle production after QGP formation are separated. The former is obtained from nuclear modifications at the initial state taking into consideration the shadowing of gluon distribution proposed previously to describe the $p_T$ spectra in $pp$ collisions. These modifications were incorporated by the multiple scattering formalism in the color dipole picture. The second part of the spectra considers effects of plasma formation until the freeze-out at the final state and has been parametrized by BGBW distribution having parameters fitted from ALICE data. We verified that the inclusion of shadowing  introduces modifications in the fitted parameters, especially in more peripheral collisions where the nuclear effects can be mostly explained by the modifications of gluon distribution at the initial state.

\section*{Acknowledgments}

This work was financed by the Brazilian funding
agency CNPq.

\bibliographystyle{h-physrev}
\bibliography{referenciasAA}

\end{document}